\documentclass[12pt]{iopart}

\usepackage{graphicx}
\usepackage{dcolumn}
\begin{document}

\title{Thermal conductivity of the diamond-chain compound Cu$_3$(CO$_3$)$_2$(OH)$_2$}

\author{J. C. Wu$^1$, J. D. Song$^1$, Z. Y. Zhao$^1$, J. Shi$^2$, H. S. Xu$^1$, J. Y. Zhao$^1$, X. G. Liu$^1$, X. Zhao$^3$ and X. F. Sun$^1$$^,$$^4$$^,$$^5$}

\address{$^1$Hefei National Laboratory for Physical Sciences at Microscale, University of Science and Technology of China, Hefei, Anhui 230026, People's Republic of China

         $^2$Department of Physics, University of Science and Technology of China, Hefei, Anhui 230026, People's Republic of China

         $^3$School of Physical Sciences, University of Science and Technology of China, Hefei, Anhui 230026, People's Republic of China

         $^4$Key Laboratory of Strongly-Coupled Quantum Matter Physics, Chinese Academy of Sciences, Hefei, Anhui 230026, People's Republic of China

         $^5$Collaborative Innovation Center of Advanced Microstructures, Nanjing, Jiangsu 210093, People's Republic of China}

\ead{xzhao@ustc.edu.cn and xfsun@ustc.edu.cn}
\vspace{10pt}
\begin{indented}
\item[]October 2015
\end{indented}

\begin{abstract}

Thermal conductivity ($\kappa$) of a distorted spin diamond-chain system, Cu$_3$(CO$_3$)$_2$(OH)$_2$, is studied at low temperatures down to 0.3 K and in magnetic fields up to 14 T. In zero field, the $\kappa(T)$ curve with heat current along the chain direction has very small magnitudes and shows a pronounced three-peak structure. The magnetic fields along and perpendicular to the chains change the $\kappa$ strongly in a way having good correspondence to the changes of magnetic specific heat in fields. The data analysis based on the Debye model for phononic thermal conductivity indicates that the heat transport is due to phonons and the three-peak structure is caused by two resonant scattering processes by the magnetic excitations. In particular, the spin excitations of the chain subsystem are strongly scattering phonons rather than transporting heat.

\end{abstract}

\pacs{66.70.-f, 75.47.-m, 75.50.-y}

{\it Keywords}: low-dimensional quantum magnets, heat transport, phonon scattering, magnetic excitations

\submitto{J. Phys.: Condens. Matt. }
%
%
%
%
%

\section{Introduction}

Low-dimensional or frustrated quantum magnets are interesting subjects of study due to their exotic magnetic properties, magnetic excitations, and quantum phase transitions (QPTs). A natural mineral azurite, Cu$_3$(CO$_3$)$_2$(OH)$_2$, has received extensive interests because of its particular magnetic structure \cite{Kikuchi1, Gu, Kikuchi2, Rule1, Kang, Aimo, Gibson, Jeschke, Rule2, Rule3, Cong1}. Cu$_3$(CO$_3$)$_2$(OH)$_2$ crystallizes in a monoclinic structure (space group $P2_1/c$) with the room-temperature lattice parameters $a =$ 5.01 \AA, $b =$ 5.85 \AA, $c =$ 10.35 \AA, and a monoclinic angle $\beta =$ 92.4$^\circ$. The Cu$^{2+}$ spins decorate the corners of diamond units forming infinite chains along the crystallographic $b$ axis, as shown in Fig. 1. It is therefore considered to be the first experimental realization of the one-dimensional (1D) distorted diamond-chain model, with the nearest exchanges $J_1 \neq J_2 \neq J_3$ (see Fig. 1).

Much experimental analysis and theoretical modeling have been performed on determining the exchange interactions, either within the 1D diamond chain or between chains. It is evidenced that the $J_2$ is the strongest interaction, leading to a separation of Cu$^{2+}$ spins into dimers and monomers \cite{Kikuchi1, Gu, Kikuchi2, Rule1, Kang, Aimo, Gibson, Jeschke, Rule2, Rule3, Cong1}. The dimerization occurs at about 20 K \cite{Kikuchi1}. The AF long-range order of the remaining monomer spins is formed at $T_N =$ 1.86 K, revealed by the nuclear magnetic resonance, electron spin resonance, and muon spin resonance studies \cite{Kikuchi1, Aimo, Gibson, Rule3}. A remarkable characteristic is the presence of a distinct 1/3 magnetization plateau at low temperatures \cite{Kikuchi1}, which was understood as a consequence of the field-induced saturation of the monomers. These results indicated that Cu$_3$(CO$_3$)$_2$(OH)$_2$ can be simply described as a combination of spin-dimer and spin-chain systems. Recently, an effective generalized spin-1/2 diamond chain model \cite{Jeschke}, with a dominant next-nearest-neighbor AF dimer coupling $J_2$, two AF nearest- and third-nearest-neighbor dimer-monomer exchanges $J_1$ and $J_3$, and a significant direct monomer-monomer exchange $J_m$ (see Fig. 1), was proposed to explain most of the experimental results.

Low-temperature heat transport has recently been found to be an effective way to probe the magnetic excitations and the field-induced quantum phase transitions \cite{Brenig, Hess1, Sologubenko1, Ando, Sun_LCO, Sun_NINO, Sologubenko2, Sologubenko3, Sun_DTN, Kohama, MCCL, Zhao_NCO, Zhao_BCVO, Niesen, Zhao_IPA, Yamashita}. In particular, a large spin thermal conductivity in spin-chain and spin-ladder systems has been theoretically predicted and experimentally confirmed in such compounds as SrCuO$_2$, Sr$_2$CuO$_3$, CaCu$_2$O$_3$, Sr$_{14}$Cu$_{24}$O$_{41}$, etc. \cite{Sologubenko4, Hess2, Hess3, Hlubek, Mohan, Karrasch}. Most of these materials have simple spin structure and strong exchange coupling, which are necessary for producing high-velocity and long-range-correlated spin excitations, while the low dimensionality strongly enhances the quantum fluctuations and ensures a large population of spin excitations. Some low-dimensional magnets however exhibit strong coupling between crystal lattice and spins, which can strongly suppress the phonon heat transport \cite{Ando, MCCL, Prasai, Sales}. In this regard, the ultrasonic, thermal expansion and neutron scattering experiments have revealed rather strong magnetoelastic couplings in Cu$_3$(CO$_3$)$_2$(OH)$_2$ \cite{Cong1, Wolff-Fabris, Cong2}. In this work, we study the heat transport properties of Cu$_3$(CO$_3$)$_2$(OH)$_2$ single crystals at low temperatures down to 0.3 K and in magnetic field up to 14 T. In zero field, the thermal conductivity ($\kappa$) has very small magnitudes and shows a three-peak structure as a function of temperature. This temperature dependence can be described by the Debye model of phonon thermal conductivity with two resonant scattering processes. The magnetic field can strongly change the $\kappa$, which has a correspondence of the change of magnetic specific heat with field. These results demonstrate that the magnetic excitations of this low-dimensional spin system are strongly scattering phonons rather than transporting heat.

\section{Experiments}

Cu$_3$(CO$_3$)$_2$(OH)$_2$ single crystals used in this work were purchased in a stone shop. The phase and the crystallinity were checked by X-ray diffraction. Using X-ray Laue photographs, the large pieces of crystals were cut into thin-plate or long-bar shaped samples with specific orientations. The thermal conductivity was measured at low temperatures down to 0.3 K and in magnetic fields up to 14 T by using a conventional steady-state technique \cite{Sun_DTN, MCCL, Zhao_BCVO, DYTO}. In these measurements, the magnetic field is either parallel to or perpendicular to the heat current ($J \rm_H$), which is along the length of the sample ($J \rm_H$ $\parallel b$). The magnetic susceptibility was measured by using a SQUID magnetometer (Quantum Design). The specific heat was measured by the relaxation method using a commercial physical property measurement system (PPMS, Quantum Design).

\section{Results}

Figure 1 shows the magnetic susceptibilities for $H \parallel b$ and $H \perp b$ ($\mu_0H =$ 1 T). The main features are: (i) each curve shows two rounded peaks at about 23 K and 5 K; (ii) the $\chi$ for $H \perp b$ is obviously larger than that for $H \parallel b$ at temperatures lower than 20 K. Apparently, the present susceptibility data  reproduce well the results in some earlier works \cite{Kikuchi1, Kang}. It has been analyzed that the peak at 23 K is related to the dominant intradimer coupling $J_2$, while the one at 5 K is due to a monomer-monomer coupling $J_m$ along the chain direction \cite{Kang}.

\begin{figure}
\centering\includegraphics[clip,width=8.0cm]{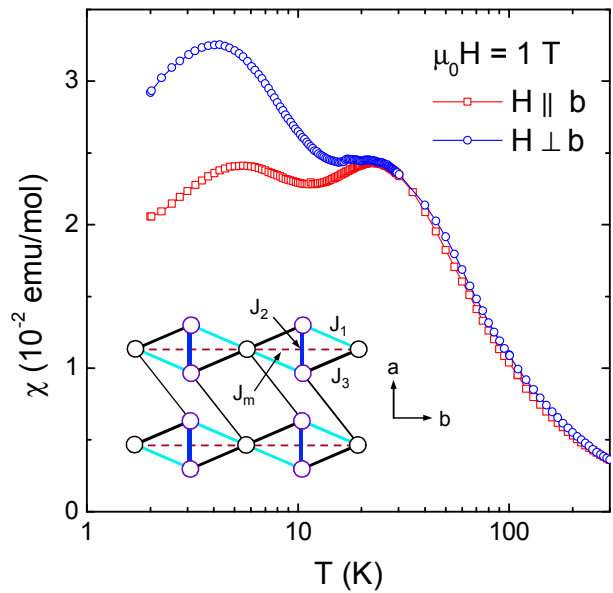}
\caption{(color online) Magnetic susceptibilities of Cu$_3$(CO$_3$)$_2$(OH)$_2$ single crystals for $H \parallel b$ and $H \perp b$ ($\mu_0H =$ 1 T). Inset: Schematic plot of the magnetic structure. There are two inequivalent  Cu$^{2+}$ ions, forming dimers (purple circles) and monomers (black circles). The exchange couplings within the diamond chain running along the $b$ include the dimer coupling $J_2$, the dimer-monomer couplings $J_1$ and $J_3$, and the monomer-monomer coupling $J_m$. The thinnest lines denote the three-dimensional couplings between diamond chains.}
\end{figure}

Figure 2(a) shows the temperature dependence of specific heat of Cu$_3$(CO$_3$)$_2$(OH)$_2$ single crystal in zero field. The data are also consistent with the earlier results \cite{Kikuchi1, Kang}. The main features are: (i) a sharp peak shows up at 1.9 K, corresponding to the long-range AF transition; (ii) a rounded peak at about 4 K; (iii) a weak hump at about 20 K. It is known that all these low-$T$ phenomena are of magnetic origin. It would be useful if the lattice specific heat can be determined. For this purpose, we measured the specific heat at temperatures up to 270 K, as shown in the inset to Fig. 2(a), which have not been studied in the earlier works. At first, a fitting to the high-$T$ data is tried by using the standard Debye formula\cite{Tari},
\begin{equation}\label{eq:eps}
C_{ph} = 3 R N \left(\frac{T}{\Theta_D}\right)^3 \int_0^{\Theta_D /T} \frac{x^4e^x}{(e^x-1)^2}\mathrm{d}x,
\end{equation}
where $x = \hbar\omega/k_BT$, $R$ is the universal gas constant, and $N$ is the total number of acoustic phonon branches. In a simplified case, $N =$ 45 (considering 15 atoms in each unit formula). However, it is easily found that the simple Debye formula cannot simulate the data at all. One known reason for the deviations of high-$T$ specific heat from the Debye model is the contribution of optical phonons at high temperatures. In such case, the Einstein model can be used for the optical phonons \cite{Svoboda, Hemberger, Janiceka}. Sometimes, different optical branches should be described with different Einstein temperatures. In the present work, it is found that the high-$T$ data can be fitted by the formula
\begin{equation}\label{eq:eps}
\fl C_{ph} = \ 3N_D R \left(\frac{T}{\Theta_D}\right)^3 \int_0^{\Theta_D /T} \frac{x^4e^x}{(e^x-1)^2}\mathrm{d}x+\sum_{i=1}^3 N_{Ei} R \left(\frac{\Theta_{Ei}}{T}\right)^2 \frac{\mathrm{exp}(\Theta_{Ei}/T)}{[\mathrm{exp}(\Theta_{Ei}/T)-1]^2}.
\end{equation}
Here, the first term is the contribution of three acoustic phonon branches using the Debye model ($N_D =$ 3), while the second term is the contributions from some optical branches. It is a simplification that the optical branches may be grouped into three ``multiple branches", which are described by different Einstein temperatures, $\Theta_{Ei}$, and different numbers of branch, $N_{Ei}$. As shown in the inset to Fig. 2(a), the calculation fits rather well to the raw data in high-$T$ range, with parameters $\Theta_D =$ 215 K, $\Theta_{E1} =$ 141 K, $\Theta_{E2} =$ 361 K, $\Theta_{E3} =$ 899 K, $N_{E1} =$ 6, $N_{E2} =$ 14, and $N_{E3} =$ 17. In passing, it should be noted that the anharmonic corrections of the phonon specific heat are not considered in the above analysis \cite{Svoboda}.

Subtracting this lattice specific heat from the raw data, one can get the low-$T$ magnetic specific heat, as shown in Fig. 2 (b). It becomes clear that, besides the sharp peak at the N\'eel temperature, the magnetic specific heat shows two broad peaks centering at about 4 and 20 K. Using the same model describing the magnetic susceptibility, these two features of the magnetic specific heat can be attributed to the contributions of the $S =$ 1/2 Heisenberg AF chain (monomers with the intrachain exchange $J_m$) and the $S =$ 1/2 AF dimer (with $J_2$), respectively \cite{Kang}. The numerical results of the magnetic specific heat of the $S =$ 1/2 Heisenberg AF chain has been known for years \cite{Bonner, Jongh, Xiang}. At not very low temperatures, it can be approximately expressed with a series expansions,
\begin{equation}\label{eq:eps}
C_{c} = \frac{3R}{16} \left(\frac{J_c}{T}\right)^2 \biggl[1 + \sum_{i=1}^\infty d_n \left(\frac{J_c}{T}\right)^n \biggr]^{-1},
\end{equation}
where $J_c$ is the spin exchange in the unit of kelvin. The coefficients $d_n$ are
\begin{equation}\label{eq:eps}
d_1 = -\frac{1}{2}, \ d_2 = \frac{9}{16}, \ d_3 = -\frac{1}{8}, \ d_4 = \frac{7}{128}, \ \cdots.
\end{equation}
The magnetic specific heat of the isolated $S =$ 1/2 AF dimers has a Schottky-like expression \cite{Jongh, Hiroi, Vasiliev},
\begin{equation}\label{eq:eps}
C_{d} = 3R \left(\frac{J_d}{T}\right)^2 \frac{\mathrm{exp}(-J_d/T)}{[1 + 3\mathrm{exp}(-J_d/T)]^2},
\end{equation}
where $J_d$ is the intra-dimer spin exchange in the unit of kelvin. With Eqs. (3) and (5), the magnetic specific heat of Cu$_3$(CO$_3$)$_2$(OH)$_2$ can be written as,
\begin{equation}\label{eq:eps}
\fl C_{m} =\frac{3 n_c R}{16} \left(\frac{J_m}{T}\right)^2 \biggl[1 + \sum_{i=1}^\infty d_n \left(\frac{J_m}{T}\right)^n \biggr]^{-1} \
+ \frac{3 n_d R}{2} \left(\frac{J_2}{T}\right)^2 \frac{\mathrm{exp}(-J_2/T)}{[1 + 3\mathrm{exp}(-J_2/T)]^2},
\end{equation}
where $n_c$ and $n_d$ are the numbers of chain spins and dimerized spins per formula unit, respectively. The experimental data of $C_m$ with $T >$ 3 K can be fitted rather well using Eq. (6), with the parameters $J_m =$ 8.14 K, $J_2 =$ 58.0 K, $n_c =$ 1.16, and $n_d =$ 2.48, as shown in Fig. 2(b). The values of $J_m$ and $J_2$ are quite close to those in the earlier studies. The values of $n_c$ and $n_d$ are a bit larger than the expected values of 1 and 2. The fail of fitting at $T <$ 3 K is due to the AF transition that is not described by the above formulas.

\begin{figure}
\centering\includegraphics[clip,width=7cm]{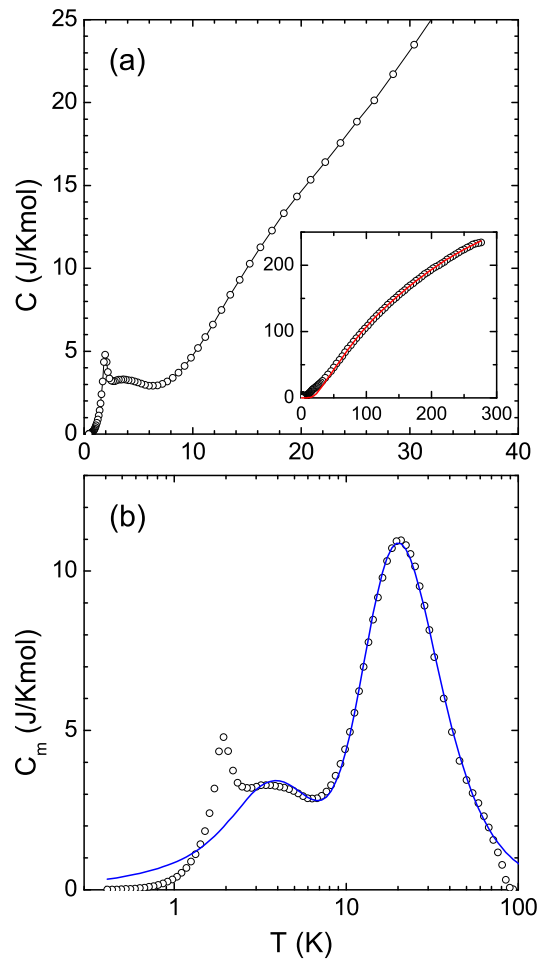}
\caption{(Color online) (a) Temperature dependence of the low-temperature specific heat of Cu$_3$(CO$_3$)$_2$(OH)$_2$ single crystal in zero field. Inset: the specific-heat data in a broader temperature range. The solid line (red) shows the fitting using the modified formula (2) of the lattice specific heat, with the consideration of optical phonons. (b) The low-$T$ magnetic specific heat obtained by subtracting the calculated phonon contribution from the raw data. The solid line shows the fitting to the data at $T >$ 3 K by using formula (6).}
\end{figure}

\begin{figure}
\centering\includegraphics[clip,width=8.5cm]{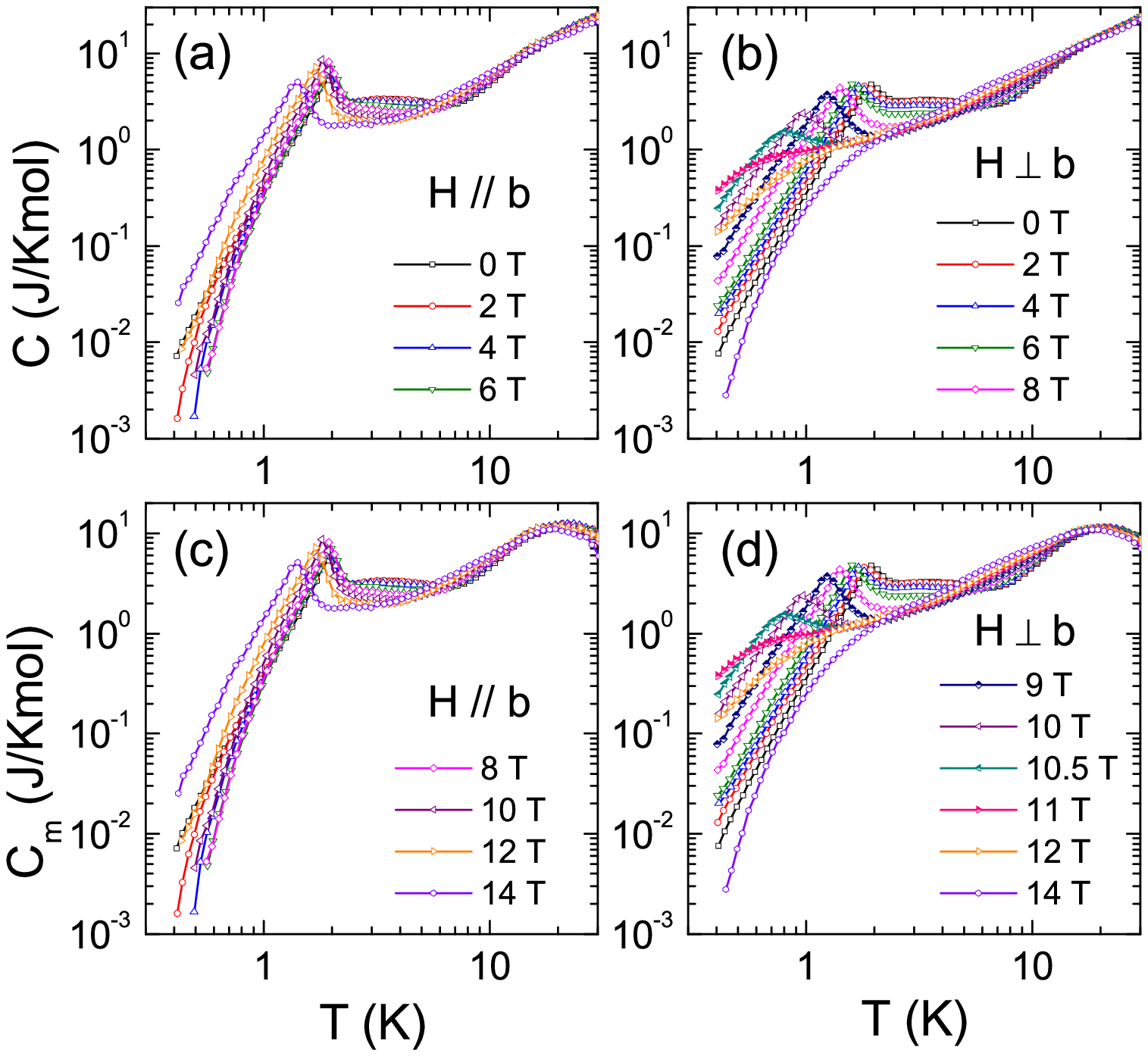}
\caption{(Color online) Temperature dependencies of low-temperature specific heat of Cu$_3$(CO$_3$)$_2$(OH)$_2$ single crystals for $H \parallel b$ (a) and $H \perp b$ (b), respectively. Panels (c) and (d) shows the corresponding magnetic specific heat $C_m$, obtained by subtracting the lattice contribution from the raw data.}
\end{figure}

The low-$T$ specific heat of Cu$_3$(CO$_3$)$_2$(OH)$_2$ was further measured in magnetic fields up to 14 T along or perpendicular to the $b$ axis. As shown in Figs. 3(a) and 3(b), an obvious result is that the AF transition is gradually suppressed with increasing field. It is found that for $H \parallel b$, the AF order is not suppressed by 14 T, while for $H \perp b$, the AF order is suppressed at about 11 T. This is consistent with the well-known magnetization plateau phenomenon, in which the lower critical field of the 1/3 plateau is about 16 and 11 T for $H \parallel b$ and $H \perp b$, respectively \cite{Kikuchi1}. Note that the 1/3 plateau appears when the magnetic field is strong enough to suppress the AF order of chain spins and polarize them. Furthermore, the magnetic field also changes the magnetic excitations at $T > T_N$. Subtracting the lattice specific heat from the raw data, one can get the magnetic specific heat in different fields, as shown in Figs. 3(c) and 3(d). It can be seen that the high-$T$ broad peak, caused by the spin dimers, is slightly weakened by the magnetic field. The peak position is shifted to lower temperature with increasing magnetic field, which is due to the shrinking of the gap between the singlet and triple states with Zeeman effect. In contrast, the low-$T$ broad peak, caused by the spin chains, is significantly weakened in magnetic fields. It is simply due to the suppression of magnetic excitations of spin chains.

The magnetic specific heat at $T < T_N$, which is related to the AF state, shows some peculiar behaviors with increasing field along different directions. For $H \parallel b$, the $C_m$ is gradually decreased with increasing field up to 8 T, but starts to recover in $\mu_0H >$ 8 T. In the highest field of 14 T, the $C_m$ becomes much larger than the zero-field values. However, the $C_m$ behaves rather differently for $H \perp b$: the $C_m$ is monotonically increased with increasing field up to 11 T, in which the AF order is just completely suppressed; in higher fields, the $C_m$ starts to decrease; in 14 T, the $C_m$ becomes much smaller than the zero-field values. The last phenomenon is easily understood because the magnon excitations are gapped in the spin-polarized state. The main reason for the nearly opposite changes of $C_m$ in low fields is that the spin easy axis is perpendicular to the $b$ axis, which could be concluded from the data of susceptibility and critical field \cite{Kikuchi1}.

\begin{figure}
\centering\includegraphics[clip,width=7cm]{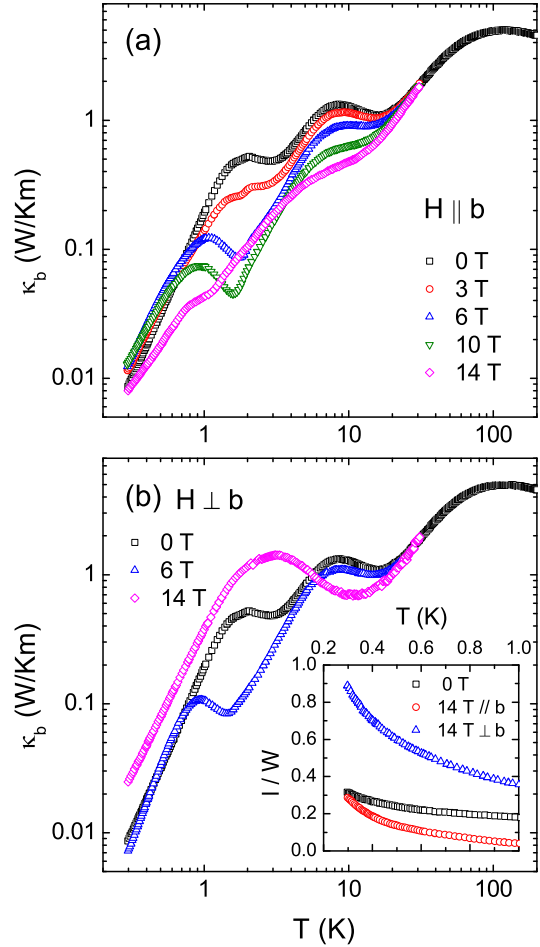}
\caption{(Color online) The $b$-axis thermal conductivity of  Cu$_3$(CO$_3$)$_2$(OH)$_2$ single crystal in zero field and several magnetic fields for $H \parallel b$ (a) and $H \perp b$ (b). Inset: the temperature dependencies of the phonon mean free path $l$ divided by the averaged sample width $W$ for the cases of zero field and 14 T $\parallel b$ or $\perp b$.}
\end{figure}

Figure 4 shows the temperature dependencies of the $b$-axis thermal conductivity of of Cu$_3$(CO$_3$)$_2$(OH)$_2$ single crystal in zero field and several magnetic fields for $H \parallel b$ and $H \perp b$. The most remarkable result is that the magnitudes of $\kappa$ are very small. It is known that single crystals of most insulators can exhibit good phonon conductivity at low temperatures, having $\kappa$ values at least one order of magnitude larger than that of Cu$_3$(CO$_3$)$_2$(OH)$_2$ \cite{Berman, Ziman}. Furthermore, in zero field the temperature dependence of $\kappa$ is very complicated. A peculiar feature of the $\kappa(T)$ is a three-peak structure at about 2, 8 and 100 K. The positions of these peaks are rather different from that of phonon peak in insulators, which usually locates at 10--20 K. Considering the small values of $\kappa$, these peaks are actually caused by the presence of two valley-like minimums at 3 and 17 K. In general, the possible reasons of these minimums in $\kappa(T)$ curves of the magnetic materials could be either the strong phonon scattering by critical spin fluctuations at some magnetic phase transitions or the phonon resonant scattering by some magnetic impurities or lattice defects \cite{Ando, MCCL, Zhao_NCO, Sologubenko4, Prasai, Berman, Ziman, Sun_GBCO, Wang_HMO, Sologubenko5, Hofmann}. It has been known that Cu$_3$(CO$_3$)$_2$(OH)$_2$ does not exhibit any magnetic phase transition at 3 and 17 K, as the magnetic susceptibility and specific heat data indicate. Therefore, the phonon resonant scattering is likely relevant. In this regard, the magnetic excitations, evidenced as two broad peaks at 4 and 20 K of the magnetic specific heat, can be the resources of phonon scattering. As already discussed these two peaks are related to magnetic excitations of spin chains (monomers) and spin dimers. It is notable that at the N\'eel transition of 1.9 K the $\kappa(T)$ shows a only small kink-like feature. It looks like that the magnon excitations of the long-range ordered state contribute weakly to the heat transport. However, the effects of magnetic field on the low-$T$ $\kappa(T)$ indicate that there are rather strong phonon scattering by the magnons (see below).

Also shown in Fig. 4 are the $\kappa(T)$ curves in several magnetic fields up to 14 T along or perpendicular to the $b$ axis. It can be seen that the directions of magnetic field play an important role in changing the thermal conductivity. In magnetic fields along the $b$ axis, the higher-$T$ valley at 17 K gradually becomes shallower, accompanied with the suppression of the 8 K peak. This behavior is reminiscent of that found in other low-dimensional spin systems with spin dimerization, such as CuGeO$_3$ and SrCu$_2$(BO$_3$)$_2$ \cite{Ando, Hofmann}. It is easily understood that the higher-$T$ valley is caused by the resonant scattering of phonons, due to the spin gap separating the dimerized singlet ground-state and the excited triplet states. On the other hand, the lower-$T$ feature of $\kappa(T)$ changes with field in a different way. In 3 T $\parallel b$, the weak kink related to the AF transition becomes a bit more pronounced (at 1.8 K), and the 3 K valley becomes shallower. In higher fields of 6 and 10 T, these two features disappear, while a strong ``dip" appears at about 1.7 and 1.6 K, respectively. Increasing field further to 14 T, the dip is weakened and evolutes into a kink- or shoulder-like feature at about 1.1 K. Since these characteristic temperatures are rather close to the AF transition temperatures probed by specific heat (see Fig. 3), it is likely that the feature of 6--14 T data are related to the magnetic transitions. The disappearance of the 3 K valley with increasing field seems to correspond with the suppression of spin-chain excitations. Moreover, the $\kappa$ at subkelvin temperatures changes non-monotonically with the field; that is, the $\kappa$ increases with field up to 10 T but decreases much in 14 T.

For $H \perp b$, the $\kappa(T)$ in 6 T displays similar behavior as that in 6 and 10 T along the $b$. That is, a dip appears at about 1.6 K, a temperature close to the AF transition in this field. However, in 14 T $\perp b$, the $\kappa(T)$ curve is very different from the others. There is no any anomaly at $T <$ 3 K and the magnitude of $\kappa$ are significantly enhanced. Remember that for $H \perp b$, 14 T is strong enough to suppress the AF order and polarize the chain spins. Therefore, it is likely that both the magnon excitations in the AF state and the chain excitations at $T > T_N$ can scatter phonons rather strongly, and these scatterings are smeared out in high enough field.

The phonon thermal conductivity can be expressed as a kinetic formula $\kappa = \frac{1}{3}C v l$, in which $C = \beta T^3$ is the low-$T$ phononic specific heat, $v$ is the averaged sound velocity and is nearly $T$-independent at low temperatures, and $l$ is the mean free path of phonons \cite{Berman}. With decreasing temperature, the microscopic scattering of phonons are gradually smeared out and the $l$ increases continuously until it reaches the averaged sample width $W = 2\sqrt{A/\pi}$, where $A$ is the cross-section area of sample \cite{Berman}. This boundary scattering limit of phonons can be achieved only at very low temperatures and the $T$-dependence of $\kappa$ is the same as the $T^3$ law of the specific heat \cite{Berman, Ziman}. In present case, none of the low-$T$ $\kappa(T)$ curves shows the $T^3$ dependence at subkelvin temperatures. Instead, the zero-field data show a $T^{2.6}$ behavior, while those in 14 T $\perp b$ show a $T^{2.2}$ dependence. These mean that the microscopic scattering is not negligible even at temperatures as low as 0.3 K. With the $\Theta_D$ value (= 215 K) from the specific-heat data, the phonon mean free path can be calculated assuming that the $\kappa$ is purely phononic \cite{MCCL, Zhao_NCO, Sun_Comment}. The inset to Fig. 4(b) shows the temperature dependencies of the ratio $l/W$ for zero field and 14 T along or perpendicular to the $b$ axis. It is found that the $l/W$ ratio is only about 0.3 at the lowest temperatures and in zero field. In the case of 14 T $\perp b$, the $l/W$ ratio increases rather quickly with decreasing temperature and approaches 1 at the lowest temperature. These indicate that the enhancement of $\kappa$ in 14 T $\perp b$ is due to the suppression of phonon scattering.

\begin{figure}
\centering\includegraphics[clip,width=6.5cm]{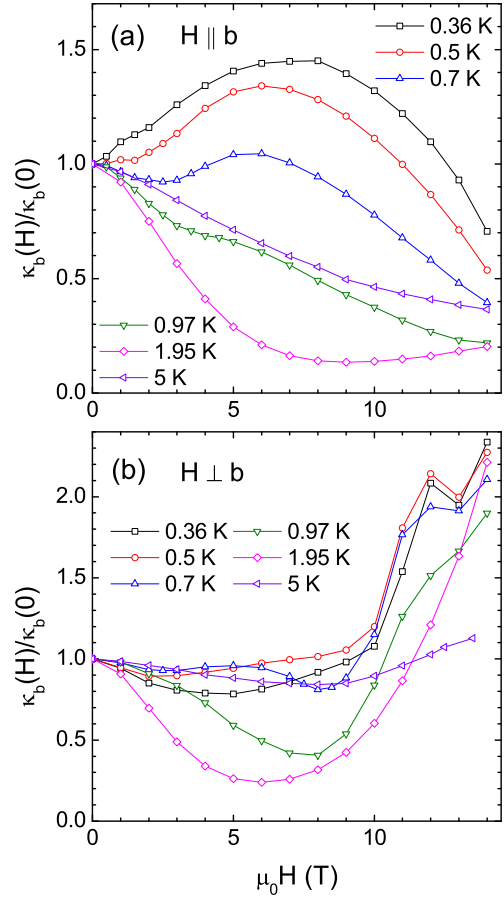}
\caption{(Color online) Magnetic-field dependencies of thermal conductivity of Cu$_3$(CO$_3$)$_2$(OH)$_2$ single crystal for $H \parallel b$ (a) and $H \perp b$ (b), respectively.}
\end{figure}

Figure 5 shows the magnetic-field dependencies of the $b$-axis thermal conductivity of Cu$_3$(CO$_3$)$_2$(OH)$_2$ single crystal for $H \parallel b$ and $H \perp b$. It confirms that the field along different directions changes the $\kappa$ in very different ways. For $H \parallel b$, the $\kappa(H)$ curve at 0.36 K shows a gradual increase with increasing $H$ up to 8 T and then a gradual decrease with further increasing $H$. With increasing temperature, there appears a low-field decrease and the 8 T maximum is weakened and moves to lower field. At 1.95 K, the maximum profile completely disappears and the $\kappa(H)$ shows a broad-valley-like behavior with the minimum locating at 9 T. At higher temperature of 5 K, the $\kappa(H)$ curve shows a continuous decrease with field up to 14 T. For $H \perp b$, the $\kappa(H)$ curves at 0.36--0.7 K (well below the zero-field $T_N$) shows a weak suppression at low field and a quick increase at about 10 T. The latter feature is likely related to the polarization of the chain spins. At higher temperatures of 0.97 and 1.95 K, the $\kappa(H)$ curves show a broad-valley-like feature accompanied with the high-field increase. At 5 K, the $\kappa(H)$ shows rather weak field dependence. It can be found that these $\kappa(H)$ behaviors have good correspondence with the field-induced changes of magnetic specific heat. Apparently, the magnetic excitations are strongly coupled with phonons and determine the field dependencies of $\kappa$ at low temperatures. In general, when the magnetic field suppresses the magnetic excitations, the phonon scattering is weakened and the $\kappa$ is enhanced. In particular, when the magnetic field is strong enough to polarize the chain spins, the low-energy magnetic excitations are strongly suppressed and the phonon scattering is significantly weakened.

\begin{figure}
\centering\includegraphics[clip,width=7.5cm]{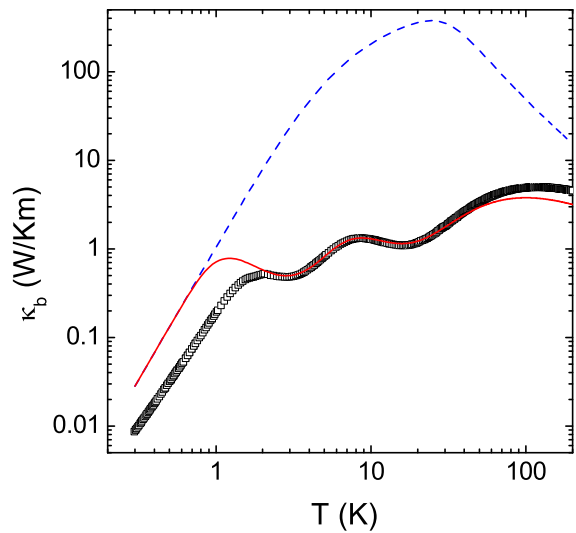}
\caption{(Color online) Comparison of the $\kappa(T)$ data in zero field and the fitting of Debye model (the solid line). The dashed line shows the calculated results with switching off the resonant scatterings.}
\end{figure}

Based on the detailed $\kappa(H)$ data, it is clear that the heat transport of Cu$_3$(CO$_3$)$_2$(OH)$_2$ is mainly phononic and there are strong magnetic scattering effects on phonons. We try a more quantitative analysis on the zero-field $\kappa(T)$ data by using a classical Debye model for the phonon thermal conductivity \cite{Berman, Ziman},
\begin{equation}\label{eq:eps}
\kappa_{ph}=\frac{k_B}{2\pi^2 v_p}\left(\frac{k_B}{\hbar}\right)^3
T^3\int_0^{\Theta_D/T} \frac{x^4e^x}{(e^x-1)^2} \tau(\omega,T)dx,
\end{equation}
in which $x = \hbar\omega/k_BT$, $\omega$ is the phonon frequency, and $\tau(\omega,T)$ is the mean lifetime of phonon. The phonon relaxation is usually defined as
\begin{equation}\label{eq:eps}
\tau^{-1} = v_p/L + A\omega^4 + BT\omega^2\exp(-\Theta_D/bT) +   \tau_{res}^{-1},
\end{equation}
where the four terms represent the phonon scattering by the grain boundary, the point defects, the phonon-phonon Umklapp scattering, and the resonant scattering, respectively. Note that the magnon scattering effect at low temperatures is not included. Two resonant scattering processes are considered for the 3 K and 17 K minimums. The higher-$T$ resonant scattering has been attributed to the (localized) singlet-triplet excitations of spin dimers. Thus, the resonant scattering rate with this process can be expressed as \cite{Sologubenko5, Hofmann}
\begin{equation}\label{eq:eps}
\tau^{-1}_{res1} = C\frac{\omega^4}{(\omega^2_1-\omega^2)^2}F(T),
\end{equation}
where $C$ is a free parameter while $F(T)$ describes the difference of thermal populations of the excited triplet and the ground singlet states. It is known that \cite{Sologubenko5, Hofmann}
\begin{equation}\label{eq:eps}
F(T) = 1 - \frac{1-\mathrm{exp}(-\Delta_1/T)}{1+3\mathrm{exp}(-\Delta_1/T)},
\end{equation}
with $\Delta_1 = \hbar\omega_1/k_BT$ the energy gap of magnetic spectrum. The lower-$T$ resonant scattering is likely related to the spin-chain excitations, since the dip of $\kappa$ and the maximum of $C_m$ are locating at almost the same temperature. Very similar phenomenon has recently found in a spin-1/2 AF chain compound CuSb$_2$O$_6$ and a two-dimensional AF material K$_2$V$_3$O$_8$ \cite{Prasai, Sales}. As firstly suggested for K$_2$V$_3$O$_8$ \cite{Sales}, this resonant scattering is likely caused by the spin excitations at the zone boundary, which have a large density of states and are corresponding to localized spin flips with zero velocity \cite{Rule3}. The resonant scattering rate associated with the spin chains can be expressed as \cite{Sologubenko4, Prasai, Toombs, Memos}
\begin{equation}\label{eq:eps}
\tau^{-1}_{res2} = D\frac{\omega^4}{(\omega^2_2-\omega^2)^2}\biggl[1 - \mathrm{tanh}^2\left(\frac{\Delta_2}{2T}\right) \biggr],
\end{equation}
with $\Delta_2 = \hbar\omega_2/k_BT$ the gap of spin excitations\cite{Sologubenko4, Prasai, Toombs, Memos}. The total resonant scattering rate is written as
\begin{equation}\label{eq:eps}
\tau^{-1}_{res} = \tau^{-1}_{res1} + \tau^{-1}_{res2}.
\end{equation}

Using formulas (7--12), the zero-field $\kappa(T)$ data are fitted. Among the parameters of this formula, $\Theta_D =$ 215 K is given by the specific heat, $v_p =$ 3850 m/s is calculated from $\Theta_D$, and $L =$ 3.8 $\times$ 10$^{-4}$ m is the sample width (the same as the parameter $W$ in above text). Figure 6 shows the best fitted result with other parameters $A =$ 1.0 $\times$ 10$^{-43}$ s$^3$, $B =$ 3.0 $\times$ 10$^{-18}$ K$^{-1}$s, $b =$ 1.5, $C =$ 2.2 $\times$ 10$^{11}$ s$^{-1}$, $D =$ 2.4 $\times$ 10$^{9}$ s$^{-1}$, $\Delta_1 =$ 50.5 K, and $\Delta_2 =$ 7.5 K. Here, the parameters $\Delta_1$ and $\Delta_2$ are comparable to the intra-dimer and intra-monomer exchanges (58.0 K and 8.14 K), respectively, determined by the specific heat. In particular, $\Delta_2$ is indeed close to the gap of $\sim$ 1.2 meV at the zone boundary detected by the neutron scattering \cite{Rule3}. Note that the fitting is not good at $T <$ 2 K, because the magnon scattering effect is not included in the calculation.

\section{Discussion and Summary}

First, it can be concluded that there is no signature that the magnetic excitations in Cu$_3$(CO$_3$)$_2$(OH)$_2$ transport heat directly. This is rather different from some well-known low-dimensional spin systems, like SrCuO$_2$, Sr$_2$CuO$_3$, CaCu$_2$O$_3$, Sr$_{14}$Cu$_{24}$O$_{41}$, La$_2$CuO$_4$, etc. \cite{Sun_LCO, Sologubenko4, Hess2, Hess3, Hlubek, Mohan, Karrasch}. The main reason is that those materials usually have much larger exchange coupling, typically being of the order of magnitude of 100 meV. The spin structure of Cu$_3$(CO$_3$)$_2$(OH)$_2$ is more complex and can be described as a combination of spin dimers and spin monomers. Although the monomers form a simple Heisenberg chain system, the intra-chain exchange ($J_m$) is quite weak. As a result, the spinon excitations of chains cannot exhibit large velocity and thermal conductivity. Instead, the magnetic excitations play a role of scattering phonons.

Second, the thermal conductivity of Cu$_3$(CO$_3$)$_2$(OH)$_2$ single crystal is much smaller than those of most insulators. The present results indicate that the phonon scattering by magnetic excitations are strong in this material. This phenomenon could be similar to those in some spin frustrated systems. A particular example is the spin-liquid Tb$_2$Ti$_2$O$_7$, in which the phonon transport is so strongly damped by the magnetic scattering that it behaves like a glassy state \cite{Li_TTO}.

Third, the magnetic scattering effect on phonons has also been signatured by the magnetostriction and magnetoelastic coupling. The elastic constant and thermal expansion coefficient of Cu$_3$(CO$_3$)$_2$(OH)$_2$ have been found to exhibit some drastic changes or anomalies at the N\'eel transition and the field-induced transitions \cite{Cong1, Wolff-Fabris, Cong2}. These have good correspondence with the particular changes of thermal conductivity. However, the heat transport data clearly indicate that apart from these drastic behaviors there are strong coupling between the spins and crystal lattice in very broad ranges of temperature and magnetic field. Based on the fitting to the zero-field $\kappa(T)$ with Debye model, one can switch off the resonant scatterings by setting $C =$ 0 and $D =$ 0. The phonon thermal conductivity obtained in this way is much larger than the experimental data, as shown in Fig. 6. As expected for resonant scattering, the damping of the phonon heat transport is most pronounced for some particular temperature ranges determined by the magnetic gaps. However, the magnetic scatterings are actually strong in a very broad temperature range.

In summary, the thermal conductivity of Cu$_3$(CO$_3$)$_2$(OH)$_2$ single crystal is studied at low temperatures down to 0.3 K and in magnetic field up to 14 T. It is found that the $\kappa$ is very small in a broad temperature range and shows quite strong field dependencies. The main conclusion is that the magnetic excitations of this quasi-one-dimensional spin system do not exhibit magnetic thermal transport, whereas they strongly scatter phonons. The temperature dependence of $\kappa$ can be described by the Debye model with the resonant phonon scatterings. Comprehensive theories quantitatively describing the scattering effect by magnons at very low temperature are called for.

\section{Acknowledgements}

This work was supported by the National Natural Science Foundation of China, the National Basic Research Program of China (Grant No. 2015CB921201), and the Fundamental Research Funds for the Central Universities (Program No. WK2030220014).

\section{Bibliography}
\label{refs}

\end{document}